\documentclass{jkas}
\usepackage{gensymb}

\def\year{2026} 
\def\volume{59} 
\def\issue{1} 
\def\beginpage{1} 
\def\received{November 12, 2025} 
\def\accepted{January 22, 2026} 
\def\published{---} 
\date{Received \received; Accepted \accepted; Published \published}

\newcommand\ion[2]{{#1}\,{\sc #2}} 
\newcommand{\msun}{\mathrm M_{\odot}}
\newcommand{\ha}{H$\alpha$}
\newcommand{\bpass}{\textsc{bpass}}

\title{%
K-DRIFT Science Theme:\\
Galactic Cirrus Clouds and Circumgalactic Medium
}

\author[1,2$\star$]{Kwang-il Seon}{0000-0001-9561-8134}
\author[1,3]{Jaehyun Lee}{0000-0002-6810-1778}
\author[1,2]{Jongwan Ko}{0000-0002-9434-5936}
\author[1]{Woowon Byun}{0000-0002-7762-7712}
\author[1,3]{Jaewon Yoo}{0000-0002-6841-8329}
\author[1]{Kyungwon Chun}{0000-0001-9544-7021}
\author[1]{Sang-Hyun Chun}{0000-0002-6154-7558}
\author[1]{Sungryong Hong}{0000-0001-9991-8222}
\author[1]{Jae-Woo Kim}{0000-0002-1710-4442}
\author[1,2]{Hong Soo Park}{0000-0002-3505-3036}
\author[1,2]{Jihye Shin}{0000-0001-5135-1693}

\affil[1]{Korea Astronomy and Space Science Institute, Daejeon 34055, Republic of Korea}
\affil[2]{Department of Astronomy and Space Science, University of Science and Technology, Korea, Daejeon 34113, Republic of Korea}
\affil[3]{Korea Institute for Advanced Study, 85 Hoegi-ro, Dongdaemun-gu, Seoul 02455, Korea}

\def\corrauthor{%
Kwang-il Seon, \email{kiseon@kasi.re.kr}
}

\def\runningauthor{%
Seon et al.
}

\def\runningtitle{%
K-DRIFT: The Circumgalactic Medium
}

\def\keywords{%
galaxies: general --- galaxies: interactions --- galaxies: stellar content --- galaxies: structure --- galaxies: dwarf galaxies --- galaxies: low surface brightness galaxies
}

\def\abstracttext{%
In this paper, we review the extended halo material and the circumgalactic medium (CGM), including both dust and gas, and discuss promising science cases that could be realized using the KASI Deep Rolling Imaging Fast Telescope (K-DRIFT). Scattered starlight from cirrus clouds in our Galaxy poses one of the major challenges to studying the low surface brightness features of extragalactic sources. Therefore, it is essential to investigate how to discriminate extragalactic sources from the cirrus cloud features. At the same time, interstellar dust clouds themselves are fundamental to understanding dust properties and the interstellar radiation field, both of which are essential for studies of chemical evolution and star formation in our Galaxy. Measuring the reddening of background sources, such as quasars, with K-DRIFT, which benefits from its broad field of view and accurate background subtraction, allows for effective detection of extended dust in galactic halos, the CGM, and intracluster space.
Observations of the \ha\ emission lines can be used to identify signatures of star formation activity within galaxies, as well as the environmental effects acting on them. Galactic winds driven by active galactic nuclei and starbursts can be traced through \ha\ emission. Strong ram pressure stripping effectively removes the interstellar medium (ISM) from galaxies. The stripped ISM becomes ionized or dissociated through mixing with the hot intracluster medium (ICM), forming \ha\ tails. The surface brightness of these \ha\ tails correlates not only with the presence of star formation in the tails but also the mixing stage of the stripped ISM and ICM. The \ha\ survey with K-DRIFT will enable the investigation of the evolutionary stages of ram pressure stripped galaxies in cluster environments, as well as the multiphase gas reservoir around galaxies and in the CGM.
}

\begin{document}
\jkashead 

\section{Introduction}
Observations indicate that galaxies are surrounded by a tenuous yet massive gaseous medium known as the circumgalactic medium (CGM). The total mass of the CGM can often exceed that found within the galaxies \citep{Tumlinson2017}. Less than 10\% of the universe’s baryons are contained in galaxies, while the remaining 90\% reside in diffuse gaseous phases, such as the CGM and the more extended intergalactic medium (IGM), which are difficult to detect due to their low densities.
The recognition of the CGM's existence originated in the mid-1950s when absorption lines of \ion{Na}{I} and \ion{Ca}{II} were discovered in the spectra of hot stars at high Galactic latitudes \citep{Spitzer1956,Munch1961}. Additionally, shortly after the discovery of absorption lines in quasars, \citet{Bahcall1969} proposed that ``most of the absorption lines observed in quasi-stellar sources with multiple absorption redshifts are caused by gas in extended halos of normal galaxies.''

The CGM\footnote{The CGM typically refers to material beyond the boundaries of galaxies but within the virial radius, as identified through optical photometry, which primarily traces starlight. In this study, however, we adopt a broader definition, using the term to describe the medium in the immediate surroundings of individual galaxies and extending beyond them. We also note that the ISM and CGM are continuously connected, making it difficult to clearly define their boundaries.} plays a critical role in the evolution of galaxies by facilitating interactions between them and the larger-scale IGM. In the 2000s, through large galaxy surveys, such as the Sloan Digital Sky Survey (SDSS; \citealt{York2000,Menard2011}) and Keck spectroscopic surveys \citep{Erb2006,Steidel2010}, as well as cosmological simulations \citep{Keres2005,Oppenheimer2006}, it became clear that addressing critical issues in understanding of galaxy evolution, including the galactic baryon deficit, the mass-metallicity relation, and quenching problems, is impossible without considering gas flows between the interstellar medium (ISM), the CGM, and the IGM \citep[see][]{Tumlinson2017}. Inflows of gas from the cosmic web are found to be necessary to sustain star formation in galaxies over cosmological timescales \citep[e.g.,][]{Keres2005,Dekel2009,Martin2019}. Simultaneously, galactic outflows are vital in regulating star formation rates (SFRs) \citep[e.g.,][]{Veilleux2005,Veilleux2020,Zhu2013}. The exchange of mass, energy, momentum, metals, and dust grains between galaxies and their surroundings drives and regulates galaxy evolution, with star formation ultimately controlled by the balance between the gas supply and the feedback that prevents it from collapsing into stars. Therefore, gas flows are of great importance for the evolution of galaxies.

The properties of the CGM have been investigated using various observational techniques, including \ha\ emission, Ly$\alpha$ absorption/emission, and metallic absorption/emission lines such as \ion{Mg}{II} and \ion{Fe}{II} \citep{Rauch1998ARA&A,Cantalupo2012,Barnes2014PASP,Zhang2016,Finley2017,Chisholm2020,Burchett2021,Leclercq2020,Leclercq2024,Langen2023}.

Ionized gas is widespread throughout the CGM and IGM. In these low-density diffuse media, a variety of sources and ionization mechanisms give rise to their characteristic multiphase structures. Emission lines originating from ionized gas---among which the Balmer H$\alpha$ line is the most commonly and easily measured---serve as tracers of photoionized gas undergoing recombination, as well as of gas that is collisionally ionized or excited and subsequently recombining and cooling.
One of the most intriguing \ha\ emission features is the ram pressure-stripped (RPS) tails of galaxies observed in galaxy clusters~\citep{gavazzi01,cortese06,cortese07,sun07,yagi07,yagi10,fumagalli14,boselli16,poggianti17,sheen17}. Because of their high peculiar velocity and high density, ram pressure stripping is particularly influential on galaxies in cluster environments. The stripped ISM is heated by the intracluster medium (ICM), forming warm ionized clouds that produce \ha\ photons. Ionizing photons emitted from young stars can also contribute to the tail \ha\ emission. \citet{lee22} showed that \ha\ filaments can primarily be attributed to collisional excitation, while the bright \ha\ cores likely arise from recombination processes in the RPS tails.

H$\alpha$ emission is also commonly observed in the galactic halos of disk galaxies, although the mechanism responsible for its excitation has yet to be fully understood \citep[see][]{Haffner2009}.
The gas component emitting H$\alpha$ halos is often referred to as extraplanar diffuse ionized gas (eDIG).  The eDIG is ubiquitous in actively star-forming galaxies \citep{Lehnert1995}. The eDIG may be ionized through various mechanisms: (1) photoionization by OB stars from star-formation regions embedded in the galactic disk or by hot low-mass evolved stars (HOLMES) in the galactic halo \citep{Reynolds1984,Flores-Fajardo2011}, (2) photoionized turbulent mixing layer models \citep{Binette2009}, (3) cooling photoionized gas \citep{Dong2011}, and (4) dust-scattered H$\alpha$ emission originating from \ion{H}{II} regions \citep{Witt2010,Dong2011,Seon2012}. Morphologically, eDIG could be composed of three components: (1) a diffuse ionized gas with a typical vertical extent of a few kpc, (2) approximately kpc-scale fine structures, such as filaments or bubbles, at the disk-halo interface, and (3) large-scale ($\gtrsim$10 kpc) filamentary structures connecting the galaxy to its environment.

In addition to the gas components, there is also compelling evidence for the presence of dust outside galactic disks or in galactic halos.
Characterizing the extent, quantity, and chemical composition of dust in the CGM is essential for evaluating its metal enrichment and for understanding the mechanisms driving material outflows from galaxies, which in turn influence galaxy evolution.
\citet{Seon2014} and \citet{Hodges-Kluck2014} found an extended dust disk layer in edge-on galaxies using far-ultraviolet (FUV) observations \citep[see also][]{Shinn2015,Hodges-Kluck2016,Jo2018,Shinn2018,Shinn2019,Boettcher2024}. Extraplanar dust in edge-on galaxies has also been observed through the analyses of far-infrared (FIR) and submm observations of its thermal emission \citep[e.g.,][]{Mosenkov2022}. In optical observations, although it is challenging to directly detect dust in the CGM, composite spectra of quasi-stellar objects (QSOs) from the SDSS have enabled the detection of dust in intervening absorption systems towards QSOs \citep{York2000,York2006}.

In this paper, we aim to review studies that are directly or indirectly relevant to halos and the CGM, exploring the potential for observation using the Korea Astronomy and Space Science Institute (KASI) Deep Rolling Imaging Fast Telescope (K-DRIFT; \citealt{Ko2026}). The K-DRIFT telescopes are designed with a unique optical configuration and employ an observing strategy optimized for studies of the low-surface-brightness (LSB) objects. The initial prototype (K-DRIFT Pathfinder) was tested through on-sky observations from 2021 to 2022. The main model, `K-DRIFT Generation 1 (G1),’ has been successfully completed and has demonstrated considerable improvements over the prototype. It employs an off-axis freeform three-mirror design with a 300\,mm aperture, a $ 4.43 \degree \times 4.43 \degree$ field of view, an $f/3.5$ focal ratio, and a pixel scale of $1.96''$. This design enables accurate and efficient detection of LSB features with flat-fielding errors lower than 0.1\% by minimizing stray light and reducing the wings of the point-spread function (PSF) in the image plane.

The remainder of this paper is organized as follows. In Section 2, we begin by discussing cirrus clouds, which must be investigated to distinguish LSB extragalactic features from foreground dust-scattered light. We also examine dust located in galactic halos and around galaxies, as well as dust in the larger-scale environment of galaxy clusters and in the CGM.
Gas components in the CGM are discussed in Section 3, with a focus on H$\alpha$ emission from jellyfish galaxies and the widespread H$\alpha$ emission in and around galaxies, which trace the multiphase nature of the gas.
Section 4 concludes with additional subjects related to the CGM, which may be investigated through K-DRIFT observations. A brief summary is provided in Section 5.

\begin{figure}
\centering 
\includegraphics[width=0.45\textwidth]{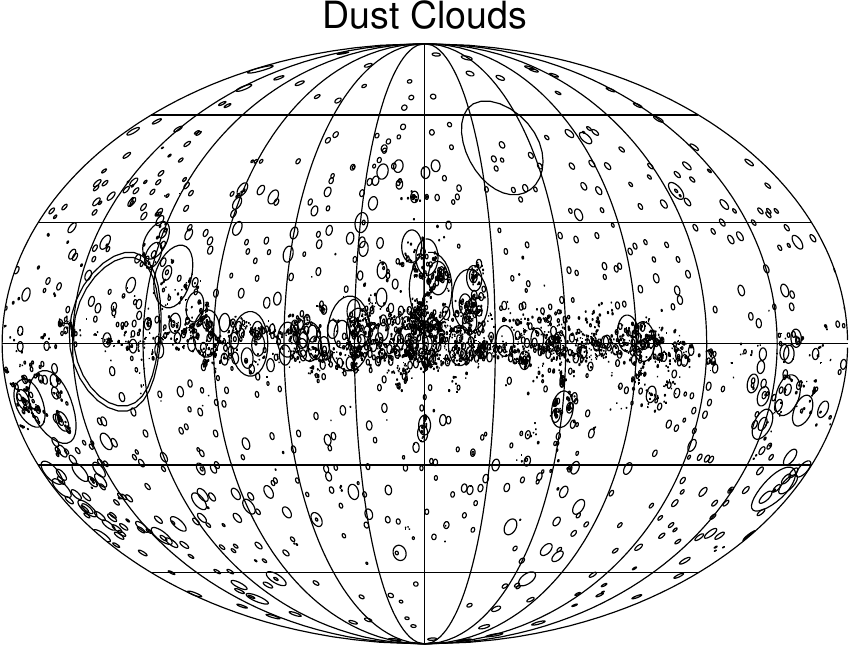}
\caption{Molleview projection map showing the locations and angular sizes of cirrus clouds in Galactic coordinates, with coordinate grids marked at 30$^\circ$ intervals. The data is taken from the catalog of dust clouds compiled by \citet{Dutra2002}, revealing a total of 5004 cirrus clouds. The size of the circles represents the rough, angular sizes of individual clouds.}
\label{fig:cirri_catalog}
\end{figure}

\begin{figure*}
\centering
\includegraphics[width=0.7\textwidth]{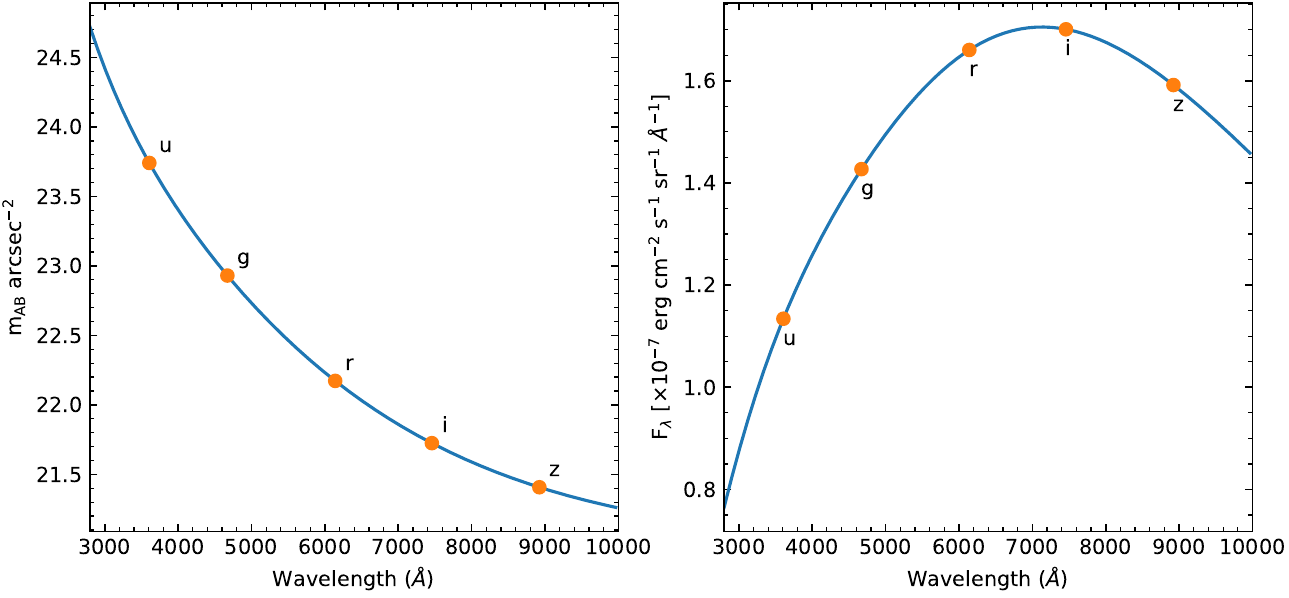}
\caption{Interstellar radiation field of \cite{Mathis1983_ISRF} and \cite{Draine2011_book}. AB magnitudes (left panel) and fluxes (right) in the SDSS $u$, $g$, $r$, $i$, and $z$ bands are shown as orange circles.}
\label{fig:ISRF}
\end{figure*}

\section{Dust}

\subsection{Cirrus Clouds in Our Galaxy}

Modern deep optical surveys allow us to investigate key aspects of the hierarchical evolution of galaxies. These aspects are closely related to LSB features in the universe, such as stellar halos and tidal features surrounding galaxies \citep{Bullock2005,Duc2015,Fliri2016}, as well as the intracluster light (ICL) in galaxy clusters \citep{Uson1991,Mihos2005,Montes2014,Montes2018}.

Despite the advancements in deep optical surveys, studying the faint outskirt regions of galaxies is a challenge due to the presence of interstellar material in the Milky Way. These clumpy and filamentary dust clouds, called ``Galactic cirri,'' scatter starlight in optical, near-infrared (NIR), and ultraviolet (UV) wavelengths, making it difficult to distinguish them from extragalactic objects. This obstacle complicates the examination of LSB extragalactic features, even at high Galactic latitudes. Figure~\ref{fig:cirri_catalog} shows the locations and sizes of the 5004 known cirrus clouds \citep{Dutra2002}, with the size of the round shapes representing the typical angular sizes of individual clouds. The figure shows only the identified cirrus clouds. It is likely that many more diffuse, thus not easily identifiable, dust clouds are present in our Galaxy than are shown in the figure.

\begin{figure*}[h]
\centering
\includegraphics[width=0.9\textwidth]{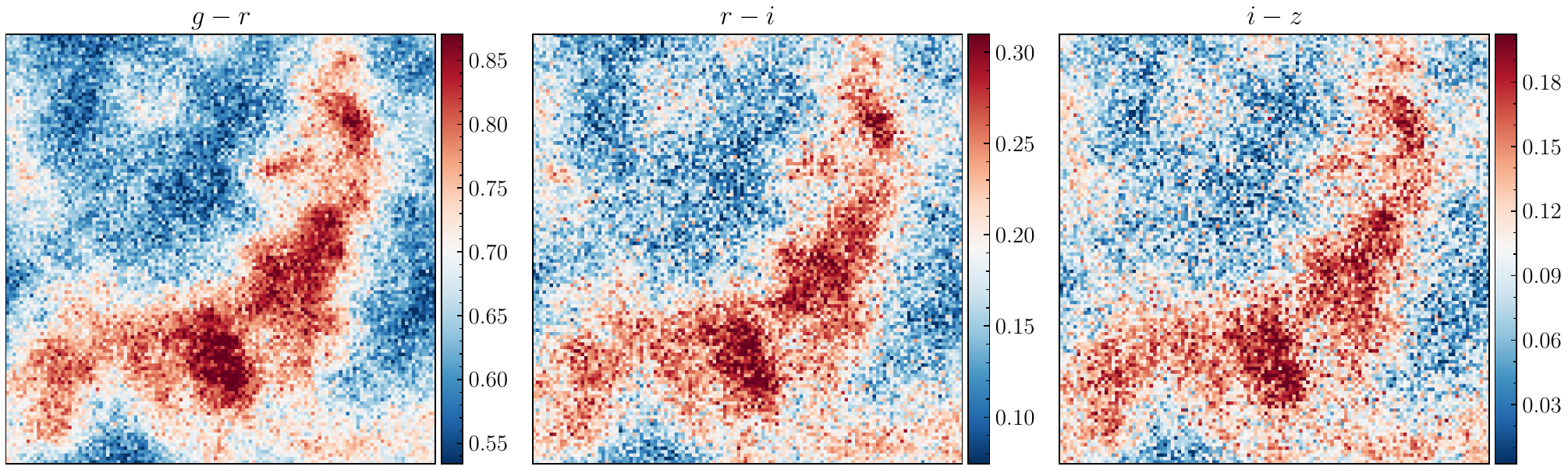}
\bigskip{}\medskip{}
\includegraphics[width=0.9\textwidth]{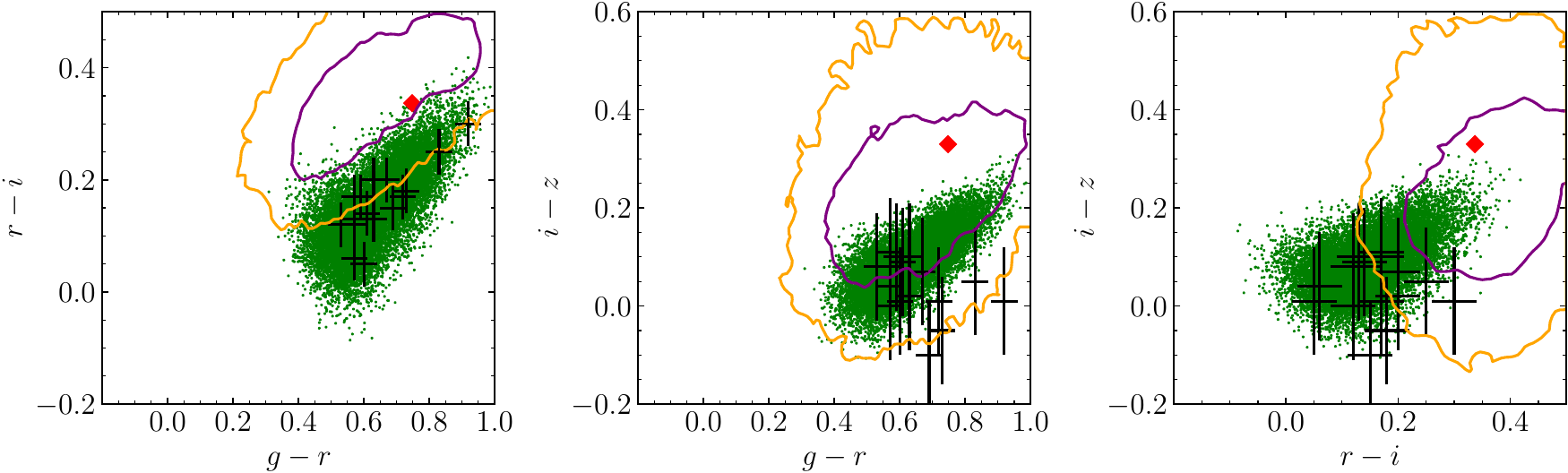}

\caption{Color maps and color-color diagrams. Top: Color maps obtained from Monte-Carlo simulations. Bottom: Color-color diagrams for the models. In the bottom panels, simulation results are shown in green dots. The black crosses represent the observational results from \cite{roman2020}. The red diamonds denote the original colors of the ISRF of our Galaxy \citep{Mathis1983_ISRF,Draine2011_book}. The purple and orange contours denote the 1$\sigma$ and 2$\sigma$ significance levels for the colors measured from LSB galaxies in the Stripe82 regions \citep{roman2017,roman2020}. In the model calculations, the $i$- and $z$-band fluxes of the ISRF were lowered by a factor of 1.1 to match the observed color-color diagrams.}
\label{fig:cirrus_model}
\end{figure*}

\begin{figure}[h]
\centering
\includegraphics[width=0.45\textwidth]{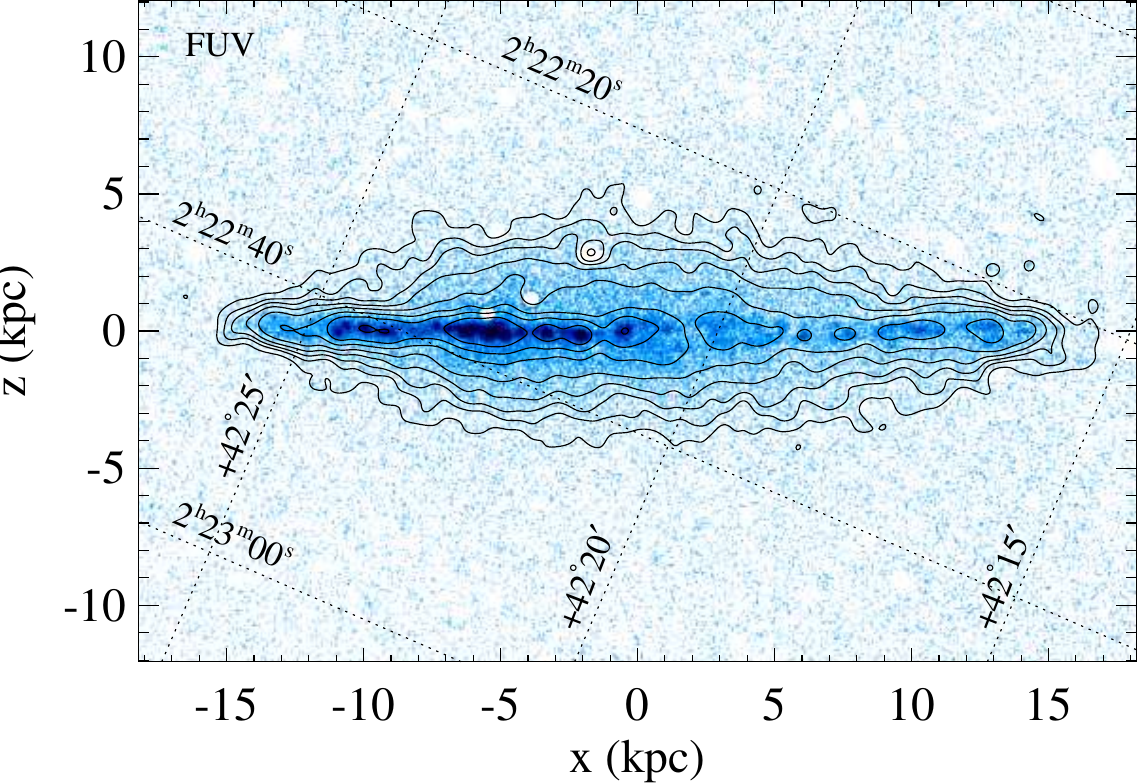}
\bigskip{}\medskip{}
\includegraphics[width=0.45\textwidth]{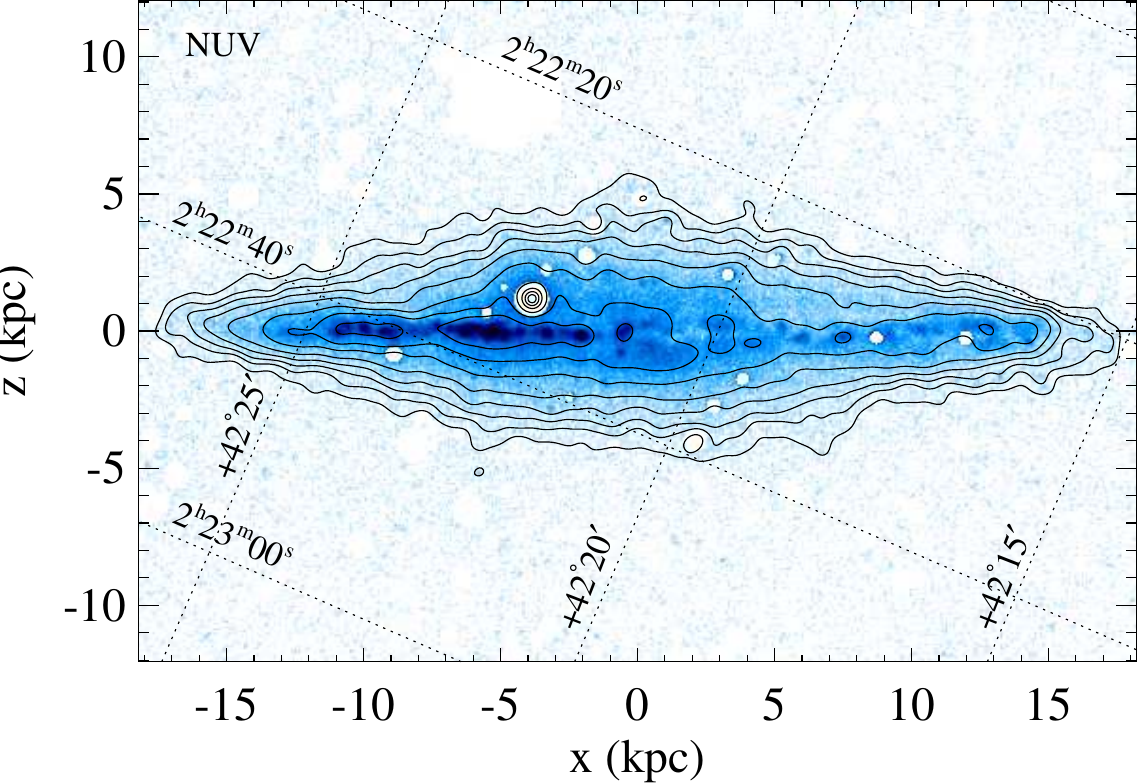}

\caption{GALEX FUV and NUV maps of NGC 891. The images were rotated such that the major axis of the disk is horizontal. Contour levels are $I=250, 450, 700, 1000, 1500, 2500, 4000$, and 7000 photons cm$^{-2}$ s$^{-1}$ sr$^{-1}$ \AA$^{-1}$. Concentric contours and white regions are artifacts due to the masking of foreground stars. The abscissa ($x$) and ordinate ($z$) denote the physical coordinates along the major and minor axes, respectively, in units of kpc, calculated assuming a distance of 9.5 Mpc to NGC 891. Right ascension and declination coordinates are also indicated by dotted lines. Adopted from \citet{Seon2014}.}
\label{fig:GALEX_NGC891}
\end{figure}

FIR full-sky maps taken by the IR Astronomical Satellite (IRAS) mission or the Planck Space Observatory are commonly used to identify cirrus clouds. However, due to their poor spatial resolution, these maps are not very effective for studying small angular scale features of extragalactic objects \citep{roman2020}. The Herschel Space Observatory provides relatively higher spatial resolution FIR data (FWHM $\sim 18^{\prime\prime}$ in the 250 $\mu$m band), but it only covers a modest portion of the sky. Therefore, it has been suggested that using high-resolution optical data itself to identify cirrus clouds is more effective. \citet{roman2020} explored the optical properties of cirrus using the $g$, $r$, $i$, and $z$ bands in the SDSS Stripe82 region. They found that the optical colors of the Galactic cirrus clouds tend to become redder as their 100 $\mu$m emission increases and differ significantly from those of extragalactic sources, exhibiting a bluer $r-i$ color for a given $g-r$ color. This behavior results from the wavelength dependence of the scattering cross-section and enables the detection of cirrus clouds through simple color relations. \citet{Smirnov2023} developed techniques based on machine learning and neural networks to isolate cirrus filaments from foreground and background sources in SDSS Stripe 82 data, utilizing the distinctive colors in the optical bands. They concluded that most filaments exhibit specific color ranges of $0.55\lesssim g-r \lesssim 0.74$ and $0.01\lesssim r-i \lesssim 0.33$.

The optical colors of cirrus clouds primarily depend on the spectral shape of the interstellar radiation field (ISRF) and the wavelength dependence of the dust scattering cross-section.
In order to understand the colors of cirrus clouds and how they differ from those of external galaxies, we developed a dust radiative transfer model \citep{Seon2025}. We assume that the isotropic radiation field with the spectral shape of Figure \ref{fig:ISRF} is incident upon the surfaces of the medium.
The figure shows the spectral energy distribution (SED) of the ISRF of the solar neighborhood, as initially defined by \cite{Mathis1983_ISRF} and later adjusted by \cite{Draine2011_book} to improve agreement with the COBE-DIRBE photometry. The left and right panels show the ISRF in units of magnitude per arcsec$^{-2}$ and intensity, respectively. The dust density field is created using a fractal algorithm to mimic the lognormal density distribution of the ISM, based on the algorithms developed by \cite{Seon2014} and \cite{SeonDraine2016}. Dust radiative transfer calculations were performed using the code \texttt{MoCafe}, developed by \cite{Seon2015} and \cite{SeonDraine2016}.

Figure \ref{fig:cirrus_model} presents the main results of the Monte Carlo radiative transfer models applied to a sample dust density field.
The top three panels show maps for the $g-r$, $r-i$, and $i-z$ colors from a model calculation, while the bottom panels show color-color diagrams between these three colors. In optically thick lines of sight, the clouds appear to be relatively red compared to the surrounding thin regions. In the color-color diagrams, the green dots represent the simulated data, expected to be observed by a distant observer from the dust cloud. The crosses denote the observational data obtained from \cite{roman2020}, and the red diamonds represent the colors of the input ISRF SED. In this calculation, the intensities in the $i$ and $z$ bands were reduced by a factor of 0.9 to better match the data. It is clear that the simulation results reproduce the observational data well, even though our dust cloud models are only a statistical representation of the density distribution of the observed cirrus clouds. The figure shows that the colors of the scattered light are bluer than those of the incident stellar radiation. Therefore, cirrus clouds can be distinguished from external LSB features if extragalactic objects have similar colors to or are redder than the stars in our Galaxy.

The results demonstrate that the difference in colors between cirrus clouds and LSB galaxies is primarily due to the fact that scattering by dust leads to bluer colors, as shorter wavelengths are more easily scattered by dust grains. The results also indicate that multi-band observations are essential for distinguishing between cirrus clouds and extragalactic LSB features when using optical data alone. More specifically, extragalactic features have redder $r-i$ colors than cirrus clouds at a given $g-r$ color. We also note that the tidal features of gas-rich galaxies exhibit bluer colors than their host galaxies, owing to star-formation triggered in tidal or stream features \citep{Schombert1990,Neff2005}, whereas those of gas-poor galaxies in cluster environments show no significant color change \citep{Pippert2025}. Figure 12 of \cite{roman2020} shows the $r-i$ versus $g-r$ color tracks for stellar populations with ages ranging from 0.1 to 12 Gyr and various metallicities, demonstrating that the $r-i$ color at a given $g-r$  color is always redder than that of cirrus clouds. Therefore, tidal features of gas-rich galaxies are expected to have redder $r-i$ colors than cirrus clouds at the same $g-r$ color. This result indicates that tidal features can be distinguished from cirrus clouds in the color–color diagram. Consequently, at least three bandpasses, or equivalently two-color observations, are required for the K-DRIFT mission to reliably separate extragalactic LSB features from cirrus features in crowded regions where complex foreground cirrus clouds are expected.

Additionally, it is suggested that the UV bandpass would be beneficial for differentiating extragalactic LSB features from Galactic cirrus clouds, as dust scattering is more pronounced in the UV than at optical wavelengths. \citet{Seon2011} demonstrated that scattered light becomes stronger at wavelengths with higher scattering albedo, indicating that colors combining UV and optical wavelengths may exhibit a clearer distinction.

\subsection{Extended Halo Dust around Galaxies}
Dust in spiral galaxies is predominantly concentrated within a thin disk layer with a scale height of $\approx$0.1--0.2 kpc. However, many attempts have been made to detect `extraplanar' dust (i.e., dust residing outside the galactic plane) by investigating edge-on disk galaxies, across multiple wavelengths, including optical, UV, NIR, mid-infrared (MIR), and FIR \citep[e.g.,][]{Greenberg1987,Howk1997,Alton2000,Rossa2004,Thompson2004,Bianchi2011,Hodges-Kluck2014,Seon2014, Bocchio2016,Yoon2021,Chastenet2024}.
The extraplanar dust features are attributable to the expulsion of dust from the galactic plane via stellar radiation pressure and/or (magneto)hydrodynamic flows, such as galactic fountains and chimneys. Therefore, their investigation is essential for understanding the galactic baryonic cycle.

In early studies, filamentary dust structures were observed up to a few kpc above the galactic plane in nearby edge-on spiral galaxies using high-resolution optical images \citep[e.g.,][]{Howk1997,Thompson2004}. These filamentary features were traced in absorption against the background starlight, suggesting the presence of relatively `dense' dust clumps that are visible only to heights limited by the vertical extent of the background starlight.
In an attempt to search for diffuse extraplanar dust in galactic halos, \citet{Seon2014} found evidence of an extended diffuse dust layer, based on the fact that dust grains at high altitudes should appear as a faint, extended reflection nebula illuminated by starlight \cite[see also][]{Hodges-Kluck2014, Seon2018, Shinn2018}. The scattered light could be clearly distinguished from direct starlight only when the scale height of the light source is lower than that of the extraplanar dust. Therefore, FUV or near-ultraviolet (NUV) observations of edge-on galaxies are best for detecting extraplanar dust, because OB stars are known to have a scale height of $\lesssim$0.2 kpc and lack a bulge or halo component.
Figure \ref{fig:GALEX_NGC891} shows the FUV and NUV images of the well-known edge-on galaxy NGC 891, which demonstrate that the emissions are extended not only along the major axis, but also along the minor axis due to dust scattering.
In addition, \citet{Seon2018} demonstrated that the optical ($V$, $R$, and $B$ bands) polarization maps observed in edge-on galaxies \citep{Fendt1996,Scarrott1996a,Scarrott1996b} could not be explained without assuming the presence of a diffuse dust layer, thus providing strong evidence supporting the existence of the halo dust component.

More direct evidence for the presence of vertically extended dust has also been obtained from MIR dust emission observed with Spitzer and FIR dust emission observed with Herschel, which trace small and large dust grains, respectively \citep{Bianchi2011,Bocchio2016,Yoon2021}. Recently, the James Webb Space Telescope (JWST) also detected dust emission in NGC 891 out to $\sim$4 kpc from the disk, in the form of filaments, arcs, and superbubbles at NIR and MIR wavelengths \citep{Chastenet2024}.

We note that cirrus clouds in our Galaxy are, in general, known to be located relatively close to the Galactic plane. Unlike line features, dust emission or absorption features provide no velocity information.
Therefore, the direct detection of high-altitude dust in our Galaxy would be impossible due to contamination from low-altitude dust along the line of sight. However, the depletion of gas-phase elements into dust grains has been reported in high velocity clouds (HVCs)---including Complex C and the Smith Cloud---infalling onto the Milky Way \citep{Cashman2023,fox2023,Vazquez2025}.
The detection of dust provides an important constraint on the origin of HVCs, as dust grains indicate that the gas has been processed through galaxies rather than being purely extragalactic. This result reinforces the widespread existence of halo dust in disk galaxies, including our Galaxy.

The direct detection of halo dust in edge-on galaxies with K-DRIFT may be hindered by confusion between direct starlight and scattered light, unless a UV bandpass is employed. However, it would be possible to investigate them indirectly. \citet{Jo2018} compared the vertical profiles of extraplanar H$\alpha$ emission with those of UV emission for 38 nearby edge-on late-type galaxies and found that the detection of `diffuse' extraplanar dust, traced by vertically extended scattered UV starlight, always coincides with the presence of extraplanar H$\alpha$ emission. They also found a strong correlation between the scale heights of the extraplanar H$\alpha$ and UV emissions; the scale height of H$\alpha$ was found to be lower than that of FUV. These findings suggest the multiphase nature of the halo material located above the galactic plane and thus provide an indirect method for studying the multiphase medium in galactic halos using a narrow-bandpass filter centered on H$\alpha$, which is planned to be installed in K-DRIFT in the future \citep{Ko2026}.

Using optical broadband observations alone is less effective for measuring extraplanar dust compared to UV and H$\alpha$ observations. However, when combined with radiative transfer models and multi-wavelength observations, such as FIR, optical broadband data can still play a valuable role in tracing the vertical extent and properties of dust in disk galaxies \citep{Xilouris1999, Bianchi2007, DeGeyter2015, Seon2018}. Investigating whether the inclusion of extraplanar dust---alongside the geometrically thin dust layer with a scale height of $\sim$200 pc---improves the agreement of radiative transfer models with deep observations of edge-on galaxies, taken with the K-DRIFT telescope optimized for detecting LSB features, would be worthwhile. Moreover, deep optical images of edge-on galaxies from the K-DRIFT survey can also reveal filamentary absorption features produced by optically thick, clumpy dust clouds---structures thought to arise from galactic fountains---across a much larger sample of galaxies. This deep imaging survey of edge-on galaxies may enable a statistical study of the correlation between the amount of filamentary dust (and hence galactic fountain activity) and host galaxy properties, such as the SFR.

\subsection{Dust in the Intracluster and Intergalactic Media}
The presence of heavy elements and dust in the ICM and  the CGM may represent both direct evidence for, and a consequence of, the ejection of galactic material.
The existence of dust in the ICM has long been a subject of investigation and remains unresolved. Contrary to the common assumption that the hot ICM prevents dust from surviving, numerous but limited observations either directly indicate the presence of dust in the intracluster space or suggest its existence as being consistent with observational data \citep[see][]{Polikarpova2017,Shchekinov2022}. 
This dust component in the ICM and IGM could significantly affects the heating and cooling of both media. \citet{Montier2004} found that dust IR emission can be the dominant cooling mechanism of the ICM when its temperature exceeds $T=10^7$ K and the dust-to-gas mass ratio surpasses $2\times 10^{-5}$. On the other hand, when the temperature is sufficiently low ($T\lesssim 10^5$ K) and the UV background is strong enough, dust grains can act as an efficient heating agent for the IGM via the photoelectric effect. These two opposite effects operate on different spatial scales: cooling dominates on large scales (in the ICM), while heating is more efficient on small scales (near quasars or star-forming regions). Thus, the dust component in the ICM/IGM may be critical to its thermal state, and consequently, to the structure formation of the universe.

\citet{Zwicky1951} was the first to propose the existence of dust in galaxy clusters by comparing the number of distant galaxy clusters located behind the rich cluster Coma with those observed in its vicinity.
\citet{Karachentsev1969} confirmed this result with a similar estimate of $A_{\rm V}\sim 0.2$--0.3 mag for a sample of 15 clusters, including Coma.
\citet{Bogart1973} and \citet{Boyle1988} found an anti-correlation between the positions of high-redshift QSOs and low-redshift galaxies, which can be explained by a model in which dust in foreground galaxy clusters obscures QSOs located behind them. In their studies, the $B$-band extinction in the clusters of the Abell catalog was estimated to be $A_{\rm B} \sim 0.2$ mag. This value corresponds to 10\% of the dust-to-gas mass ratio in the Milky Way. \citet{Romani1992} found that optically selected quasars appear to avoid foreground rich clusters, showing a considerable deficit in number. This finding allowed them to estimate an average sample extinction of $A_{\rm V} \sim 0.4$ mag, with significant clumpiness (up to $\Delta A_{\rm V} \sim 1$ mag).

However, some subsequent observations indicated that only a substantially smaller amount of dust is present in the diffuse gas of galaxy clusters. A study by \citet{Maoz1995}, using the $V-I$ color distribution of radio-selected quasars behind rich Abell clusters, yielded only upper limits on the reddening, with $E(B-V)<0.05$ mag. \citet{Stickel2002} analyzed Infrared Space Observatory observations of six Abell clusters at 120 and 180 $\mu$m and detected marginal dust emission in Abell 1656 (Coma), corresponding to a dust extinction of $A_{\rm V}\ll 0.1$ mag and a dust-to-gas mass ratio of $\sim$$10^{-6}$, but found no evidence of extinction in other five clusters. \citet{Chelouche2007} reported an $E(B-V)$ of a few $\times 10^{-3}$ for sight lines passing within $\sim$1 Mpc of the centers of $z\sim 0.2$ galaxy clusters. The measured reddening was found to exhibit a trend of flattening within the core and to decrease slowly outside in an inversely proportional manner to the distance from the center. The inferred dust-to-gas mass ratio is less than $5\%$ of the value in the local ISM of the Milky Way. \citet{McGee2010} found a similar trend from a sample of 70,000 low-redshift groups and clusters in the SDSS. These results can be interpreted as being due to the rapid destruction of dust in the hot environment of central regions of clusters and the partial survival in less hostile conditions at the periphery. More recently, \citet{Longobardi2020} found evidence of diffuse dust in the ICM of the Virgo cluster, with $E(B-V)\simeq 0.042$ mag and extinction $A_{\rm V}\simeq 0.14$ mag, assuming the Large Magellanic Cloud extinction law. The inferred dust-to-gas mass ratio is nearly 20 times lower than that of the Milky Way, indicating that dust in hot intracluster plasma undergoes destruction, primarily through thermal sputtering.

The existence of circumgalactic dust on Mpc scales was demonstrated by \citet{Mernard_Scranton2010}, who detected a smooth extinction profile extending from 20 kpc to several Mpc using the gravitational magnification and dust reddening of 85,000 background quasars by 24 million galaxies at $z\sim 0.3$ derived from the SDSS. They found its projected density to follow a distribution similar to mass profile $\Sigma_{\rm dust}\sim r_{\rm p}^{-0.8}$. A similar measurement was performed by \citet{Peek2015}, who used a background sample of luminous red galaxies, whose small color dispersion enabled precise measurement of dust halo reddening with a smaller sample size than that used by \citet{Mernard_Scranton2010}.

Detecting IR emission from individual galaxy clusters is challenging due to the extremely low level of this emission, along with fluctuations in the IR sky, Galactic cirrus clouds, and background galaxies. Instead, stacked analyses of IR emission from hot intracluster dust have been conducted. \citet{Montier2005} detected statistically significant IR emission toward galaxy clusters at 12, 25, 60, and 100 $\mu$m by co-adding IRAS maps for a total of 11,507 galaxy clusters. By complementing the IRAS data with observations from the Planck satellite, which ranges from 100 to 857 GHz, \citet{PlanckCollaboration2016} provided new constraints on the IR spectrum of thermal dust emission in galaxy clusters using a stacking approach applied to several hundred objects from the Planck cluster sample. They confirmed the existence of dust in galaxy clusters at low and intermediate redshifts, yielding an SED that resembles that of the Milky Way. They also propose that galactic sources---particularly spiral galaxies---are the primary contributors to the cluster dust. However, the majority of gas-phase metals in galaxy clusters are believed to be produced by elliptical and S0 galaxies.

Minimizing photometric uncertainties is essential for studying reddening in the ICM and IGM. Systematic uncertainties are the primary challenge in achieving this minimization. For instance, efforts to reach deeper surface brightness levels are hindered by local background fluctuations, introduced by a combination of zodiacal light, moonlight, stray light from stars, PSF, Galactic cirrus, and other factors \citep[see][]{Knapen2017book, Holwerda2021book}. These sources of contamination can introduce inaccuracies in the calibration process, such as in sky background determination, resulting in significant photometric uncertainties in the LSB regime. To address these challenges, K-DRIFT adopted an off-axis design with a small focal ratio to enhance sensitivity to faint light and reduce the systematic uncertainties introduced by the optical system. As a result, K-DRIFT is expected to offer a promising opportunity for the systematic investigation of dust in the ICM and IGM by taking advantage of its large field of view and optimized background subtraction capabilities.

\begin{figure*}
\centering 
\includegraphics[width=0.95\textwidth]{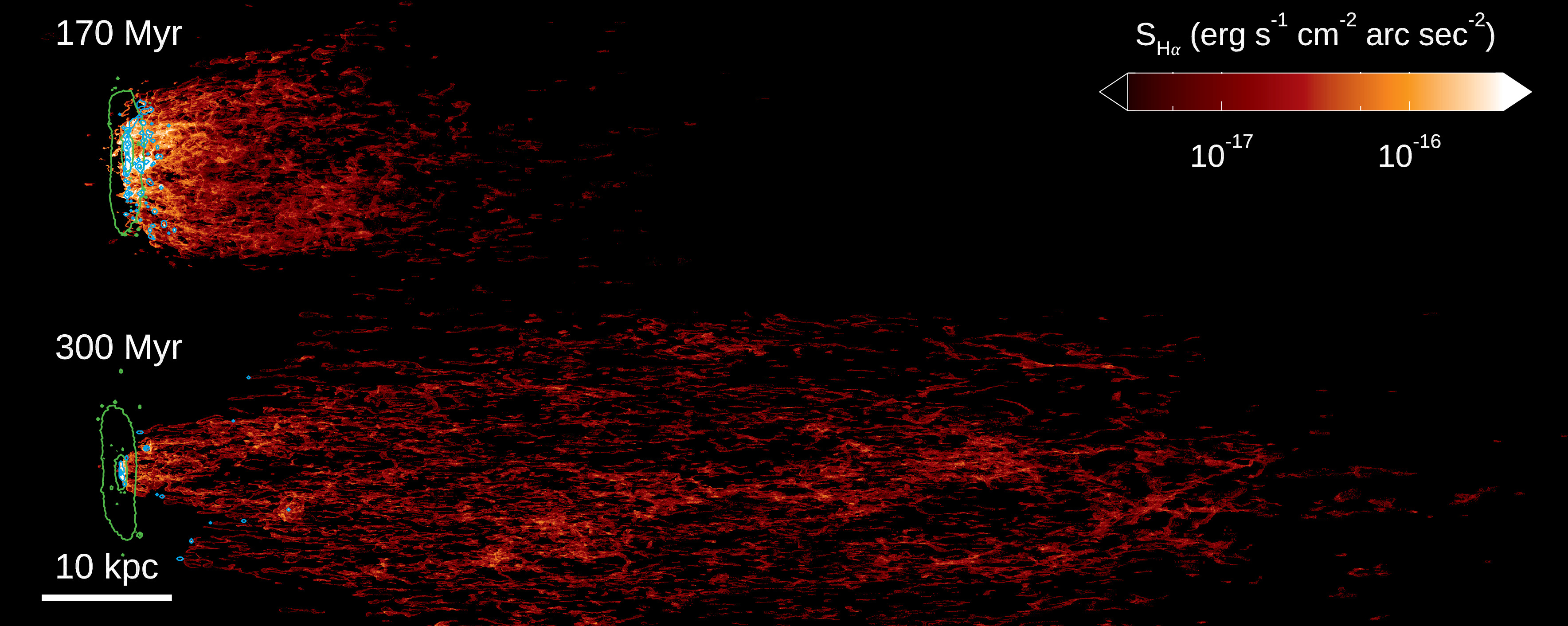}
\caption{\ha\ emission of a simulated RPS galaxy after encountering a strong ICM wind \citep{lee26}. The green and blue contours denote the distribution of all stars and of stars younger than 20\,Myr, respectively. Numerous star-forming clumps are visible in the near tail at 300\,Myr. The star-forming regions coincide with bright \ha\ cores, but only a small fraction of the \ha\ emission correlates with young stars in the tail.}
\label{fig:halpha_tail}
\end{figure*}

\section{Gas}

\subsection{\ha\ Emission from the Ram Pressure Stripped Gas in the Cluster Environments}

\subsubsection{Physical Origin of Jellyfish Features}
RPS is a hydrodynamical process that can directly remove the ISM from galaxies within a timescale of a few hundred Myr. Because of high peculiar velocity and medium density, RPS becomes a significant environmental effect in galaxy clusters~\citep{gunn72,davies73,boselli06,boselli22}. This process has been witnessed through jellyfish features observed in \ion{H}{i}~\citep{kenney04,oosterloo05,chung07,chung09,scott10,scott12,scott18}, CO~\citep{jachym14,jachym17,jachym19,verdugo15,lee17,lee18,moretti18}, H$\alpha$~\citep{gavazzi01,cortese06,cortese07,yagi07,yagi10,fumagalli14,boselli16,poggianti17,sheen17,yagi17}, and X-ray bands~\citep{finoguenov04,wang04,machacek05,sun05,sun06,sun10}. Some jellyfish galaxies have been observed using the integral field unit (IFU) instrument MUSE on the VLT as part of the GASP program~\citep{gullieuszik17,poggianti17,poggianti19,franchetto21}. The IFU observations enable one to find the physical origin of emission~\citep{poggianti19} and a metallicity gradient~\citep{franchetto21} in the jellyfish tails.

A few jellyfish galaxies have been intensively studied using multi-band observations of \ion{H}{i}, CO, H$\alpha$, and X-rays~\citep{jachym17,jachym19,ramatsoku25}. The presence of jellyfish features across a wide wavelength range reveals the multiphase nature of the RPS tails. It has been suggested that mixing between the stripped ISM and ICM forms the multiphase tails~\citep[see][]{sun10,sun22,franchetto21}. Since the ICM has a temperature of $T\sim10^7$--$10^8\,$K, mixing between the two media produces warm ionized clouds that emit \ha\ photons. If the warm clouds have sufficient density and size to shield themselves from the hot ICM, they can form cold clumps via radiative cooling~\citep[e.g.,][]{armillotta16,armillotta17,gronke18}.

Many recent efforts have been made to examine the multiphase nature of RPS tails using hydrodynamical simulations. For example, \citet{lee20,lee22} investigated the impact of varying ram pressure on dwarf galaxies with different gas abundances using a set of radiation hydrodynamical simulations. A mild ram pressure that a galaxy experiences in the cluster outskirts effectively strips low-density clouds in the outer disk while compressing dense clouds and enhancing star formation activity within them. In contrast, strong ram pressure quickly removes the ISM from both normal and gas-rich galaxies, quenching their star formation within 300\,Myr. In the simulations conducted by \citet{lee20,lee22}, prominent jellyfish features with clumpy clouds and star formation in the tails emerge only when a gas-rich dwarf galaxy experiences strong ram pressure. In this case, a large amount of the ISM is stripped and mixed with the ICM, forming warm ionized clouds that can cool and condense into molecular clumps within a few hundred Myr. This process causes the metallicity gradient in the cold clumps along the tail, which is observed in the GASP project~\citep{franchetto21}. Consequently, the stellar populations in the tails exhibit lower metallicity with increasing distance from the disk.

\subsubsection{\ha\ Emission from Jellyfish Galaxies}

\ha\ emission from the RPS tails provides direct evidence of the ionization of the stripped ISM~\citep[e.g.,][]{yagi10,yagi17}. \citet{sun22} claimed that the surface brightness ratio between \ha\ and X-rays in the RPS tails suggests the role of mixing between the ICM and stripped ISM. Bright \ha\ clumps are believed to be induced by young stars formed in the RPS tails~\citep[e.g.,][]{jachym17,sheen17,jachym19,poggianti19}. Thus, \ha\ emission provides plenty of information about the evolutionary stage of the RPS galaxies, particularly when combined with the information obtained from other wavelengths.

To examine the \ha\ emission in RPS galaxies, we conduct a suite of radiation magneto-hydrodynamical (RMHD) simulations using \texttt{RAMSES-RT}~\citep{rosdahl13,rosdahl15a}, an RMHD version of the adaptive mesh refinement code \texttt{RAMSES}~\citep{teyssier02,fromang06,teyssier06}. \texttt{RAMSES-RT} computes the evolution of magnetic fields using a constrained transport method~\citep{teyssier06} and traces photon flows using a moment-based radiative transfer scheme~\citep[see more details in][]{rosdahl13}. The version of \texttt{RAMSES-RT} used in the simulations computes the formation and destruction of molecular hydrogen using a modified photochemistry model, on top of the non-equilibrium chemistry and cooling of six chemical species, i.e., \ion{H}{i}, \ion{H}{ii}, \ion{He}{i}, \ion{He}{ii}, \ion{H}{iii}, and $e^-$.

We used the initial conditions of a dwarf galaxy generated by \citet{rosdahl15b} using \texttt{MAKEDISK}~\citep{springel05}. The galaxy is embedded in a dark matter halo with a mass of $M_{\rm halo}=10^{11}\,\msun$ and a virial radius of $R_{\rm vir}=89\,$kpc and its initial stellar mass is $M_\star=2.1\times10^9\,\msun$. \citet{lee22} demonstrated that a plenty stripped ISM forms prominent jellyfish features via mixing with the ICM. We thus set the initial \ion{H}{i} mass fraction to $M_{\rm HI}/M_\star=4.2$. The \ion{H}{i} disk initially has metallicity of $Z_{\rm ISM}=0.75\,Z_{\odot}$, where $Z_\odot=0.0134$~\citep{asplund09}. A magnetic field of $B_x=10^{-0.5}\mu$G is initially assumed. The galaxy is placed at the center of a cubic volume with a side length of $L=300\,$kpc. We made the galaxy evolve until it reaches a quasi-equilibrium state. After relaxation, the galaxy has $M_{\star}=3.4\times10^9\,\msun$, $M_{\rm HI}=1.9\times10^9\,\msun$, and the magnetic field strength of $|B|\sim1-10\,\mu$G in dense clouds of $n_{\rm H}>1\,{\rm cm^{-3}}$. We then imposed a wind of $v=1,000\,{\rm km\,s^{-1}}$, $n_{\rm H}=3\times10^{-4}\,{\rm cm^{-3}}$, $T=3\times10^7$ K, $Z_{\rm ICM}=0.3Z_{\odot}$, and $B_{x}=1\,\mu$G toward the galaxy in the face-on direction. We defined the moment of wind launch as $t=0\,$Myr. Given the wind parameters, it was estimated that the wind front arrives on the disk at $t\sim135\,$Myr, but the wind starts to influence the galaxy $\sim30\,$Myr earlier due to shock expansion at the wind front. The pressure of the wind ($P_{\rm ram}/k_{\rm B}=5.\times10^5\,{\rm K\,cm^{-3}}$, where $k_{\rm B}$ is the Boltzmann constant) is set to mimic the typical ram pressure that a galaxy can encounter in the cluster center~\citep[see Figure\,10 of][]{jung18}. We ignored metal enrichment by supernovae, to be able to trace the fractions of the ICM and ISM separately within each cloud. We defined the disk of the RPS galaxy as a cylindrical volume with a radius of $r=12\,$kpc, and a height of $h=\pm3\,$kpc from the galactic plane. The RPS tail was accordingly defined as the characteristic features in a long cylinder of $r=12\,$kpc and $h>3\,$kpc. More details of the simulation set will be given in \citet{lee26}.

\begin{figure}
\centering 
\includegraphics[width=0.43\textwidth]{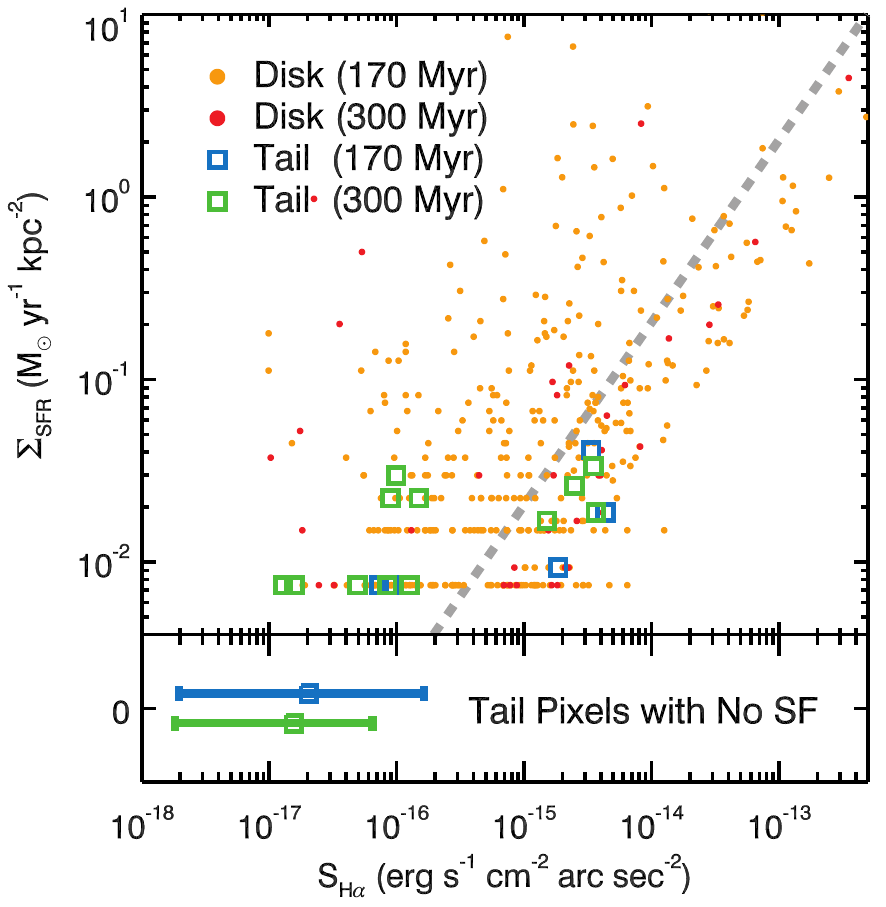}
\caption{The relation between the SFR surface density ($\Sigma_{\rm SFR}$) and $S_{\rm H\alpha}$ in the disk and tail of the RPS galaxy shown in Figure~\ref{fig:halpha_tail}, at 170~Myr and 300~Myr after the wind launch.
Open squares and filled circles represent the tail and disk regions, respectively. For clearity, all symbols are plotted in different colors. The SFR--$S_{\rm H\alpha}$ relation predicted by \bpass\ \citep{eldridge08,stanway16} is shown with a gray dashed line. The bars in the bottom panel indicate the $S_{\rm H\alpha}$ ranges of the pixels contributing to the $1^{\rm st}-99^{\rm th}$ percentile distribution of the H$\alpha$ brightness in the tail regions with no star formation at 170 Myr (blue) and 300 Myr (green). Squares on the bars denote the median values of $S_{\rm H\alpha}$.}
\label{fig:halpha_sfr}
\end{figure}

The initial setup of this simulation is identical to that of \citet{lee22}, except for the inclusion of magnetic fields. \citet{lee22} demonstrated the crucial role of mixing between the stripped ISM and ICM in the formation of molecular clumps in RPS tails. Meanwhile, it is well known that magnetic fields can stabilize the interface between two different flows by suppressing the growth of dynamical instabilities, thereby reducing the efficiency of mixing~\citep[e.g.,][]{frank96,ryu00,esquivel06,heitsch07,hamlin13,liu18,praturi19}. Indeed, \citet{lee26} showed that RPS galaxies form less fragmented and smoother tail clouds when the ICM winds are magnetized, owing to suppressed mixing, resulting in the absence of star formation in the distant tails. Interested readers are referred to \citet{lee26} for further details.

We computed the \ha\ emissivity of the simulated galaxies using the two fitting functions derived by \citet{peters17} for recombination and collisional excitation processes. Figure~\ref{fig:halpha_tail} shows the \ha\ surface brightness ($S_{\rm H\alpha}$) of an RPS galaxy at two epochs after the wind launch. In this figure, the blue contours display the regions enclosing stars younger than 20~Myr and the green contours shows the distribution of all disk stars. Numerous star-forming regions are visible in the near tail, within 10\,kpc of the galactic plane, and the majority of them coincide with local \ha\ clumps. However, most of the diffuse \ha\ emission is unrelated to the star-forming regions, indicating that physical processes other than high-energy photons are responsible for producing the majority of \ha\ photons in the tail. We examine the relation between unobscured \ha\ emission and the star formation activity of the galaxy in Figure~\ref{fig:halpha_tail}. We measured $S_{\rm H\alpha}$ and the SFR surface density ($\Sigma_{\rm SFR}$) on pixels of $73\times73\,{\rm pc}^2$ and correlated them in Figure~\ref{fig:halpha_sfr}. In the upper panel, the $\Sigma_{\rm SFR}$-$S_{\rm H\alpha}$ relation closely follows the prediction of \bpass~\citep[gray dashed line,][]{eldridge08,stanway16} in both the disk and tail, albeit with large scatter. This plot is missing some tail pixels that contain young stars but lack bright \ha\ cores, likely due to rapid decoupling driven by strong winds.

\begin{figure}
\centering 
\includegraphics[width=0.42\textwidth]{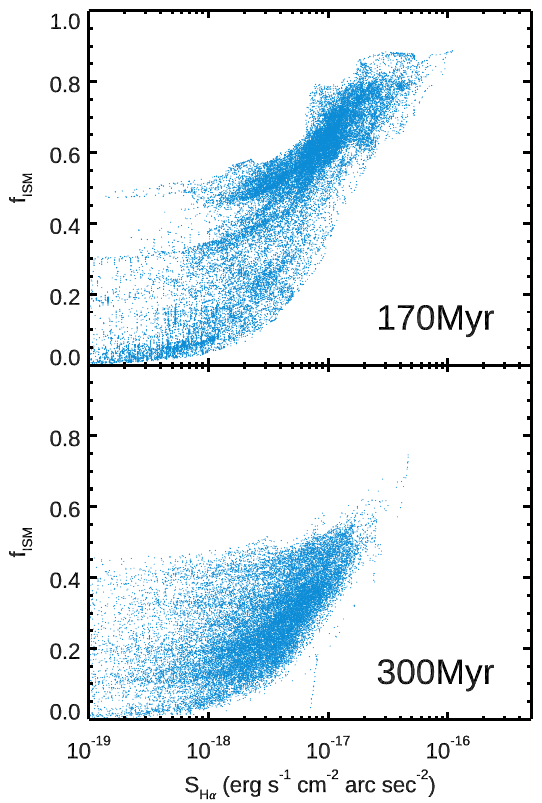}
\caption{Fraction of the ISM as a function of $S_{\rm H\alpha}$, smoothed over $1\times1\,{\rm kpc^2}$ pixels randomly sampled from the tails of the galaxies shown in Figure~\ref{fig:halpha_tail}.}
\label{fig:ha_fism}
\end{figure}

We also plotted the range of the $S_{\rm H\alpha}$ distribution covering the $1^{\rm st}$--$99^{\rm th}$ percentiles of the \ha\ brightness in the tail regions without star formation activity, in the bottom panel. The range reaches as high as $S_{\rm H\alpha}\sim2\times10^{-16}{\rm erg\,s^{-1}\,cm^{-2}\,arc\,sec^{-2}}$, as denoted by the bars. 
However, most pixels lie below the typical $\Sigma_{\rm SFR}-S_{\rm H\alpha}$ relation of star-forming regions. As mentioned in \citet{lee22}, the diffuse \ha\ emission is primarily produced by collisional excitation. Even though its surface brightness is notably lower than that of bright \ha\ cores associated with young stars, it is spatially widespread, and thus accounts for most of the tail \ha\ emission. This result suggests an $S_{\rm H\alpha}$ threshold of $S_{\rm H\alpha}\sim3\times10^{-16}\,{\rm erg\,s^{-1}\,cm^{-2}\,arc\,sec^{-2}}$, above which the emission is likely associated with star-forming regions. 

\subsubsection{Mixing of Stripped ISM with ICM}

The stripped ISM is heated by the hot ICM, forming warm ionized clouds that emit \ha\ in the RPS tails. Therefore, the brightness of \ha\ emission is expected to correlate with the amount of stripped ISM. Indeed, \citet{sun22} found a strong correlation between \ha\ and X-ray surface brightness measured over $10\times10\,{\rm kpc^2}$ regions in the tails of 17 RPS galaxies observed in the Virgo, Coma, Abell 1367, Abell 2626, and Abell 3627 clusters. The characteristic flux ratio of $F_X/F_{\rm H\alpha}\sim3$ was interpreted as the result of ISM-ICM mixing, since X-ray photons are mainly produced by hot plasma with $T\gtrsim10^7\,$K, whereas \ha\ photons mainly originate from warm ($T\sim10^4\,$K) ionized gas. Using their simulation, \citet{lee22} showed that a higher $F_X/F_{\rm H\alpha}$ ratio corresponds to a lower ISM fraction, supporting the interpretation of \citet{sun22}.

In this study, we also measure the ISM fraction and $S_{\rm H\alpha}$ in $1\times1\,{\rm kpc^2}$ pixels randomly sampled in the tails of the RPS galaxies in Figure~\ref{fig:halpha_tail} and correlate them with each other, as illustrated in Figure~\ref{fig:ha_fism}. As expected, brighter $S_{\rm H\alpha}$ values correspond to higher ISM fractions. Overall, the $f_{\rm ISM}-S_{\rm H\alpha}$ relation appears to be only weakly affected by the evolutionary stage of the RPS tail. At later epochs, the tail clouds exhibit fainter $S_{\rm H\alpha}$ and lower $f_{\rm ISM}$, but the slope of the $f_{\rm ISM}-S_{\rm H\alpha}$ relation remains nearly unchanged in log--log space. Fitting this relation yields slopes of 0.39 at 170\,Myr and 0.40 at 300\,Myr for $S_{\rm H\alpha} > 10^{-18}\,{\rm erg\,s^{-1}\,cm^{-2}\,arc\,sec^{-2}}$.

\begin{figure}
\centering 
\includegraphics[width=0.475\textwidth]{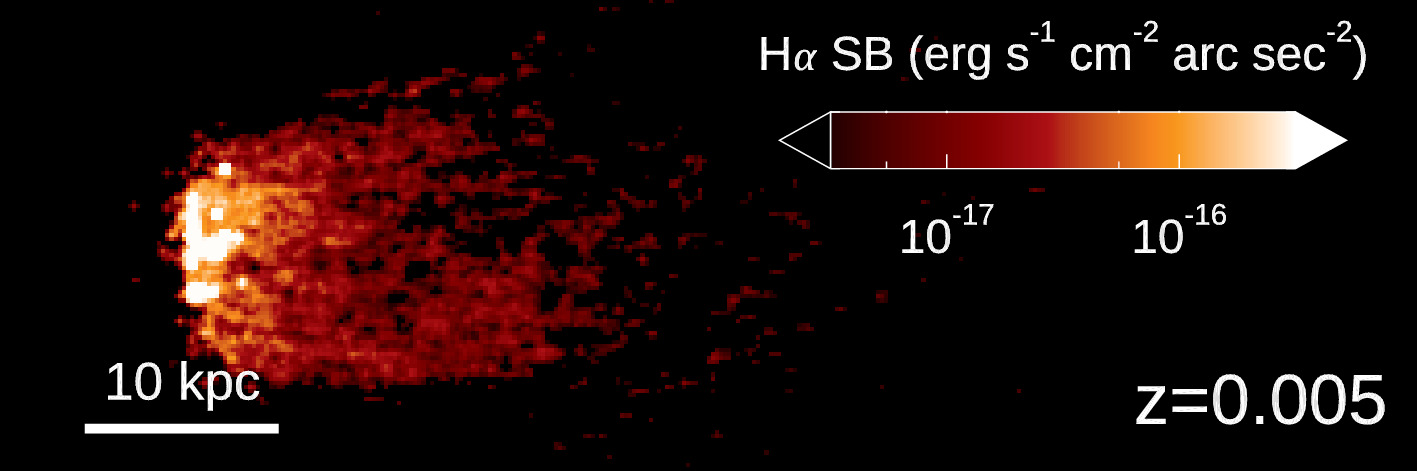}
\includegraphics[width=0.475\textwidth]{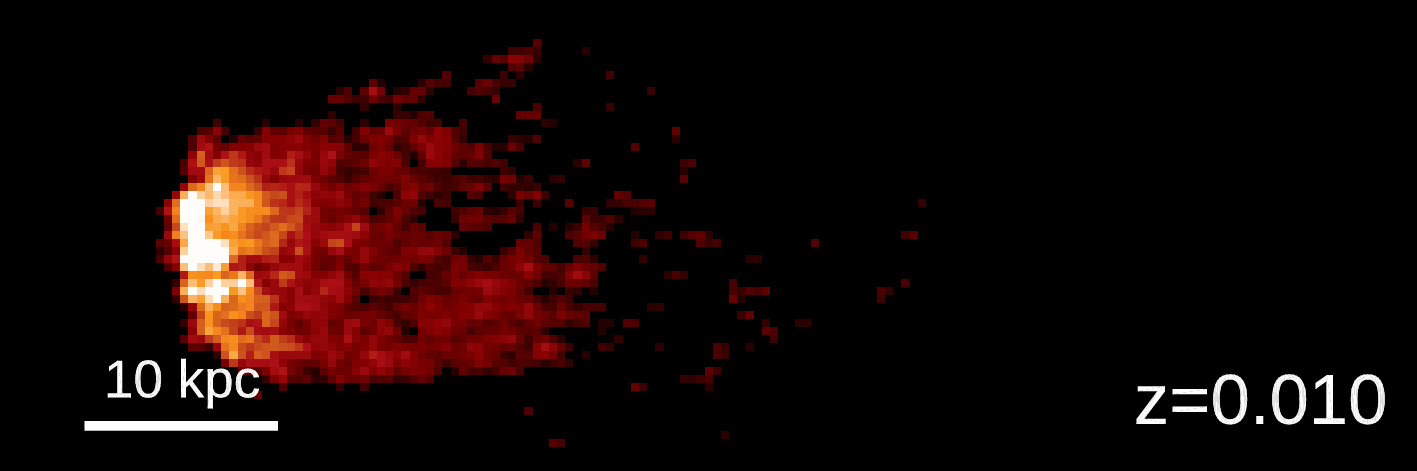}
\includegraphics[width=0.475\textwidth]{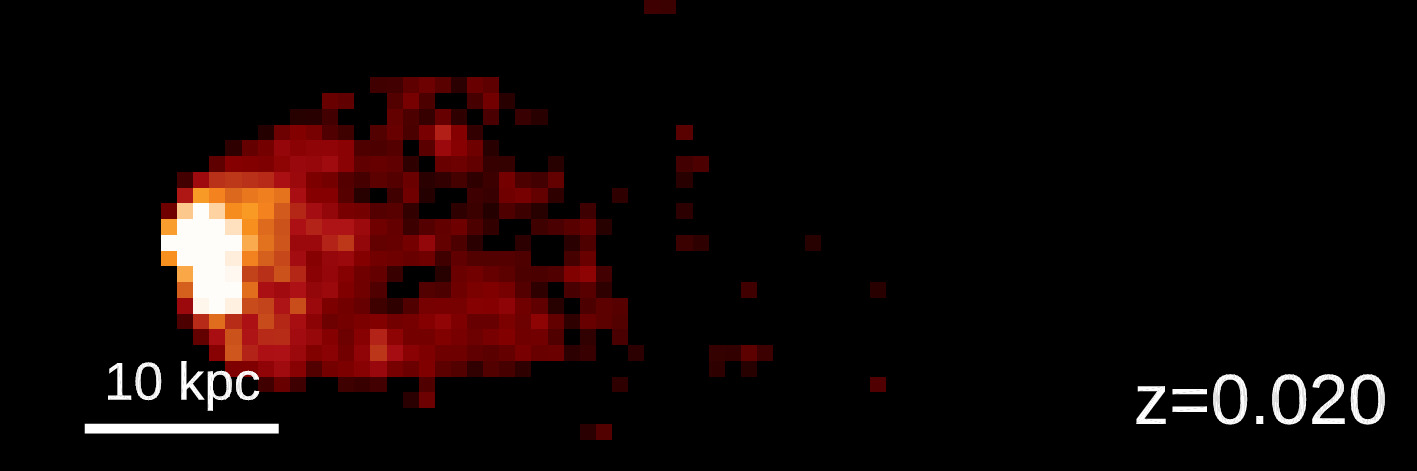}

\caption{Mock H$\alpha$ surface brightness map of the galaxy (at 170~Myr), shown in the upper panel of Figure~\ref{fig:halpha_tail}, projected onto the plane with the K-DRIFT pixel scale (2$^{\prime\prime}$) at three different redshifts.}
\label{fig:mock_ha}
\end{figure}

\subsubsection{\ha\ Observations in K-DRIFT }

One of the key topics of the K-DRIFT project is the study of diffuse light distributed throughout galaxy clusters. We have shown that \ha\ observations can provide valuable information about RPS galaxies. Here, we examine how the \ha\ tail features would appear at varying distance when observed with K-DRIFT. The main cluster targets of K-DRIFT are located at $z\sim0.005-0.02$, and the pixel scale of K-DRIFT is 2$^{\prime\prime}$. Figure~\ref{fig:mock_ha} shows mock images of H$\alpha$ brightness projected onto the plane with the K-DRIFT pixel scale. We assume three different redshifts, $z=0.005$, 0.01, and 0.02, which correspond to luminosity distances $D_{\rm L}$ (pixel scale) = 22.4~Mpc (0.107~kpc/$^{\prime\prime}$), 44.8~Mpc (0.213~kpc/$^{\prime\prime}$), and 90.3~Mpc (0.421~kpc/$^{\prime\prime}$), respectively, based on the cosmological parameters from the Planck 2018 results~\citep{Planck2020}. We applied the PSF of K-DRIFT (FWHM = 3.0$^{\prime\prime}$) to the mock image.
This mock image demonstrate that the RPS tails can be observed even at $z=0.02$ by K-DRIFT at a level of \ha\ $\sim10^{-18}\,{\rm erg\,s^{-1}\,cm^{-2}\,arc\,sec^{-2}}$ or higher.

\subsection{Extended H$\alpha$ Emission as a Tracer of the Multiphase Gas Reservoir around Galaxies and in the CGM}

Narrowband and spectroscopic observations of H$\alpha$ emission in nearby late-type, edge-on galaxies have revealed extraplanar H$\alpha$ emission in many galaxies with SFRs \citep{Rand1996,Miller2003a,Miller2003b,Rossa2003a,Rossa2003b,Ho2016,Lu2023}. Extraplanar H$\alpha$ emission is generally considered to originate from the extraplanar diffuse ionized gas (DIG). The DIG is typically understood to be sustained primarily by photoionization from ionizing photons (Lyman Continuum; LyC) produced by early-type stars in the galactic plane, which leak through the porous, clumpy ISM \citep{Reynolds1984,Haffner2009,Barnes2015}.

However, \citet{Lu2023} found that the scale height of the extraplanar diffuse H$\alpha$ emission is approximately 0.7 times that of the \ion{H}{i} 21 cm line emission \citep{Zheng2022}. This finding suggests that the extraplanar DIG is more extended than the neutral gas, because H$\alpha$ emission is proportional to the square of the electron density, whereas 21 cm emission is proportional to the density of neutral hydrogen. This, in turn, implies the existence of additional, more extended ionizing sources beyond the LyC photons leaking from the disk star-forming regions. Moreover, the H$\alpha$-to-UV flux ratio is found to be lower in the extraplanar DIG than in \ion{H}{II} regions, indicating that field OB stars residing outside star-forming regions may be important contributors to DIG ionization \citep{Hodges-Kluck2016,Jo2018}.

In addition to this extraplanar H$\alpha$ emission, the interarm regions in face-on spiral galaxies are also found to emit diffuse H$\alpha$ emission \citep{Ferguson1996,Hoopes1996,Greenwalt1998,Poetrodjojo2019}.
The diffuse H$\alpha$ emission outside bright \ion{H}{ii} regions is not only very extended but can also appear in distinct patches or filaments far from \ion{H}{II} regions. It is also well estabilished that the line ratios of [\ion{S}{ii}] $\lambda$6717/H$\alpha$ and [\ion{N}{ii}] $\lambda$6583/H$\alpha$ observed far from bright \ion{H}{ii} regions are generally higher than those within the \ion{H}{ii} regions. \citet{Sardaneta2024} examined isolated, edge-on galaxies and compared their sample with the more general sample of \citet{Rossa2003a} to study the effects of galactic structure and environment, concluding that extraplanar H$\alpha$ emission is uncorrelated with galaxy environment.

The physics of DIGs is relevant not only to the multiphase nature of the CGM but also to LyC leakage during the epoch of cosmic reionization. LyC photons can will escape through density-bounded regions or channels with low hydrogen column densities. In density-bounded regions, more LyC photons are present than hydrogen atoms can absorb; hence, the medium becomes fully ionized, and the remaining LyC photons escape once ionization is complete. In contrast, an ionization-bounded region is surrounded by a high \ion{H}{i} column density and is optically thick, making it difficult for LyC photons to escape. In massive star-forming galaxies, LyC leakage from density-bounded \ion{H}{II} regions may produce DIGs in the halos and interarm regions of the galaxies. On the other hand, LyC photons escaping from compact, low-mass galaxies at high redshift would contribute to the full ionization of the IGM. Therefore, understanding DIGs is crucial for addressing key questions in a variety of astrophysical contexts.

In addition to extraplanar H$\alpha$ emission, diffuse H$\alpha$ emission from the extended CGM, spanning tens of kpc, has also been discovered using wide-field, deep H$\alpha$ imaging techniques \citep{Watkins2018,Lokhorst2022,Xu2023ApJ}.
\citet{Watkins2018} identified an extended, diffuse H$\alpha$-emitting feature---without any embedded compact regions---located $13^\prime$ (32 kpc) north of the interacting system M51, using deep narrowband imaging observations. They attributed this feature to gas expelled from the inner regions of the M51 system, possibly due to tidal stripping or starburst/AGN-driven winds, and suggested that the gas is ionized either by shocks or a fading AGN. An H$\alpha$-emitting cloud discovered by \citet{Lokhorst2022} exhibits a shell-like morphology, with a linear extent of $0.\!\!^\circ8$ (55 kpc), and is located $0.\!\!^\circ6$ (40 kpc) northwest of M82. The cloud may consist of material lifted from M82 by tidal interactions or by its powerful starburst, or it may represent infalling gas from the cosmic web, precipitated by the superwinds of M82.
\citet{Xu2023ApJ} reported a large ionized structure in NGC 5195, extending to $\sim$10 kpc from its nucleus. This structure may be an outflow inflated by a past episode of elevated AGN activity or is associated with tidally stripped material illuminated by a luminous AGN in the past.

K-DRIFT may be able to observe these extended H$\alpha$ emissions around galaxies by combining a narrowband filter centered on H$\alpha$ with a broadband continuum filter, thereby achieving unprecedentedly LSB sensitivity compared to previous studies. Comparing H$\alpha$ observations with those in the UV and optical wavelength bands would provide critical insights into the connection between this multiphase CGM and stellar activity in galactic disks---including the effects of infalling gas from the IGM, outflows driven by AGNs and star formation, and galaxy merger processes. It is also expected that previously unknown, extended ionized structures will be serendipitously discovered through wide-field survey observations conducted with K-DRIFT.

\section{Concluding Remarks}
In this section, we briefly discuss additional topics not covered above but closely related to the subject of this paper, which are expected to provide further insights through K-DRIFT observations.

\subsection{Metallic Absorption/Emission Lines from the CGM}

Most studies of the diffuse halo gas have been conducted through absorption against bright background sources such as quasars. However, absorption-line studies along pencil-beam sightlines are generally unable to provide a complete mapping of the extended gaseous halo. Recent advances in IFU spectrograph technology have made it possible to measure spatially resolved emission lines in the CGMs of star-forming galaxies.
Pristine, metal-poor baryons fuel star formation originate from outside the halo, whereas metals are produced by stars over their lifetime. It has been found that star-forming galaxies with stellar masses spanning three orders of magnitude retain a relatively constant fraction ($\sim 20$--$25$\%) of the metals they have produced, indicating that metals are lost through outflows \citep{Tremonti2004,Peeples2014}. However, how these outflows scale with galaxy mass remain unclear. Furthermore, there have been few, if any, studies on the correlation between these metallic lines and halo dust or intracluster dust.

Neither absorption nor emission lines of metals can be observed with K-DRIFT because it lacks spectroscopic capability. Nevertheless, systematic studies of extended dust---composed of metallic elements---outside galaxies, to be carried out by K-DRIFT, could be compared with metallic-line observations. Such comparisions would, in turn, provide a more complete understanding of metallic enrichment in the CGM driven by galactic outflows.

\subsection{Relation between the CGM and ICL}
Galaxy clusters grow hierarchically through successive mergers and the accretion of smaller systems fed along cosmic filaments within large-scale structure of the universe.
This process often leaves behind an optically detectable feature known as the ICL. It is believed that the majority of the ICL originates from violent relaxation following mergers between satellite galaxies and the brightest cluster galaxy \citep{murante07}, or from tidal stripping \citep{contini14,joo25}. Meanwhile, \citet{ahvazi24}, using the TNG50 simulation, reported that a significant fraction ($\sim$10--30\%) of the ICL is formed {\em in situ}. Although star formation in outer galactic environments depends strongly on the physical prescriptions adopted in simulations, several studies have demonstrated that stars can form in RPS tails~\citep{kapferer09,lee22}. Comparing their results with previous work, \citet{lee22} pointed out that the stripping of abundant interstellar material is essential to reproduce the star formation in RPS tails. 

The correlation between the amount of material in the CGM and the fraction of the ICL may provide valuable insights into the origin of both the CGM and the ICL. K-DRIFT observations of the ICL could help investigate this correlation by combining measurements of CGM dust obtained with K-DRIFT and gas components probed by other instruments.

\section{Summary}
K-DRIFT has a unique design and observing strategy that make it well suited for observations of LSB features.
In this work, we explored the feasibility of studying the CGM and galactic outskirts with K-DRIFT, which has been developed for LSB science. To probe the CGM and halo environment, multiband observations, including H$\alpha$ narrow-band imaging and NUV data, are essential.

Distinguishing extragalactic LSB features from foreground Galactic cirrus emission is critical. To this end, K-DRIFT will investigate the properties of Galactic cirrus clouds and establish robust criteria for their identification. In particular, color-color diagrams will provide a practical diagnostic for separating extragalactic LSB objects from diffuse Galactic dust clouds.

Photometric observations of galactic halos are crucial for understanding metal enrichment in the CGM and IGM, and thus for constraining galaxy evolution. K-DRIFT observations can provide key information on relatively unexplored, extended dust component in galactic halos, both within and around galaxies, and potentially in galaxy clusters as well.
Dust in the ICM and in the CGM/IGM can be measured along lines of sight toward bright quasars and background galaxies. With its wide-area, uniformly reduced dataset, K-DRIFT will offer valuable insights into the metal enrichment history of the IGM, thereby providing important guidance for studies of galaxy evolution.

Observations of diffuse, LSB H$\alpha$ emission around galaxies are essential for investigating star formation activity and environmental processes that shape galaxy evolution.
K-DRIFT is expected to reach a surface brightness sensitivity of $\sim$$10^{-18}$ erg cm$^{-2}$ s$^{-1}$ arcsec$^{-2}$ in H$\alpha$ with only a few hours of observation. This sensitivity will enable systematic studies of H$\alpha$ halos produced by DIG in external galaxies, as well as emission associated with ram-pressure stripping in jellyfish galaxies. Narrowband H$\alpha$ observations will therefore provide a unique opportunity to conduct uniform surveys of the DIG in edge-on galaxies and to characterize ram-pressure stripping most effectively examined in jellyfish galaxies.


\acknowledgments

This research was supported by the Korea Astronomy and Space Science Institute under the R\&D program (Project No.\ 2026-1-831-03) supervised by the Korea AeroSpace Administration.
KIS is supported by a National Research Foundation of Korea (NRF) grant funded by the Korean government (MSIT; No.\  2020R1A2C1005788).
J.L. is supported by the National Research Foundation of Korea (RS-2021-NR061998). J.L. also acknowledges the support of the National Research Foundation of Korea (NRF) grant funded by the Korea government (MSIT; 2022M3K3A1093827).
This work was supported by the Center for Advanced Computation at Korea Institute for Advanced Study. 



\bibliography{references}{}





\end{document}